\documentclass[11pt]{article}

\usepackage{graphicx}
\usepackage{subfigure}
\usepackage{multirow}
\usepackage{url}

\begin{document}
\sloppy

\title{\bf Evaluating Accessible Synchronous CMC Applications}

\author{    {\bf Fernando G. Lobo}\\
            \small CENSE and DEEI-FCT\\
            \small Universidade do Algarve\\
            \small Campus de Gambelas\\
            \small 8005-139 Faro, Portugal\\
            \small fernando.lobo@gmail.com
\and
           {\bf Marielba Zacarias}\\
            \small DEEI-FCT\\
            \small Universidade do Algarve\\
            \small Campus de Gambelas\\
            \small 8005-139 Faro, Portugal\\
            \small marielba.zacarias@gmail.com
\and
           {\bf Paulo A. Condado}\\
            \small CENSE and DEEI-FCT\\
            \small Campus de Gambelas\\
            \small Universidade do Algarve\\
            \small 8005-139 Faro, Portugal\\
            \small pcondado@gmail.com
\and
           {\bf Teresa Rom\~{a}o}\\
            \small CITI-DI-FCT\\
            \small Universidade Nova de Lisboa\\
            \small Quinta da Torre\\
            \small 2829-516 Caparica, Portugal\\
            \small tir@di.fct.unl.pt
\and
           {\bf Rui Godinho}\\
            \small DEEI-FCT\\
            \small Campus de Gambelas\\
            \small Universidade do Algarve\\
            \small 8005-139 Faro, Portugal\\
            \small the.rg.space@gmail.com
\and
           {\bf Manuel Moreno}\\
            \small DEEI-FCT\\
            \small Campus de Gambelas\\
            \small Universidade do Algarve\\
            \small 8005-139 Faro, Portugal\\
            \small woreno@gmail.com           
}
\date{}
\maketitle

\begin{abstract}
This paper proposes a comprehensive evaluation methodology to measure the usability and user experience qualities of accessible synchronous computer-mediated communication applications. The method allows evaluating how the interaction between a user and a product influences the user experience of those at the other endpoint of the communication channel. 
A major contribution is given with the proposal of a user test where one of the participants tries to guess whether the other participant has a disability or not. The proposed test is inspired in the Turing Test, and is a consequence of user requirements elicited from a group of individuals with motor and speech disabilities. These ideas are tested and validated with two examples of synchronous communication applications.
\end{abstract}

\section{Introduction}
\label{sec:introduction}

Computer applications that require real-time interaction are increasingly being used, in large part due to the interconnected world that we live in.  Prominent examples include instant messaging, voice over IP applications, various sorts of online computer games and multi-user role-playing environments.
A common thing among these applications is the existence of two or more participants that communicate in a synchronous fashion. This contrasts with other kinds of computer-mediated communication (CMC) applications where the communication is asynchronous (e.g. Email, bulletin boards, newsgroups, blogs, and wikis) in which the conversation is not time-dependent and participants respond to each other when they want or have the time to do so~\cite{Sharp:07}.

Having access to these applications is very important because they contribute to a better socialization and integration in society. This is of special relevance for people with disabilities because they often feel isolated and excluded in face-to-face interactions. Information and communication technologies, and in particular, interactive applications with multiple participants, can help to attenuate their isolation. 

With respect to accessibility, a major challenge in the design of synchronous CMC applications is to reduce the time needed for people with disabilities to interact with the user interface. Addressing this problem improves the user experience on both end parties as there are no unnecessary pauses that can cause distraction and frustration among the participants involved in the communication. 

Usability is generally defined by five quality components: Learnability, efficiency, memorability, error rate and recovery, and satisfaction~\cite{Nielsen:94}. Most of these components can be quantified precisely by measuring things such as the time needed to perform a given task, the number of errors done while inputing information, and so on. The satisfaction component is more qualitative and in some circles of the HCI community tends to fall into the user experience category, where the term \emph{user experience} goes beyond usability by including aesthetics, hedonics, perceptions, emotions, judgements, and attitudes that a user has as a result of the interaction~\cite{Beauregard:07} \cite{Norman:99}.

Evaluation procedures in current CMC practice tend to be symmetric; they evaluate usability and user experience qualities resulting from the interaction between users, and do so assuming that all participants have similar capabilities. In other words, they do not regard the case of existing users with disabilities.
In this paper, we argue that such tests are not sufficient to assess the quality of the interaction in synchronous CMC applications, especially when some of the participants involved in the communication have some sort of disability which may lead to a slowdown in the communication rate. We propose that a proper evaluation for synchronous CMC should take into account the interface usability as well as the user experience qualities for all the participants involved in the communication. 

The genesis of this paper starts with the construction of a set of user requirements derived from interviews conducted with a group of  people with various degrees of motor and speech disabilities. 
One thing that stood out from the interviews was that all the individuals would like to be able to communicate in a more effective manner. Some of them even expressed that, in certain situations, they would like to communicate in such a way that their disability is unnoticed. This desire is for pragmatic reasons, not because of being ashamed of their disability.
This observation lead us to propose a test for synchronous CMC to measure if a given participant perceives whether the other participant has a disability or not. Our proposal is inspired by the Turing Test~\cite{Turing:50}, a landmark work in the field of Artificial Intelligence (AI).

Throughout the paper we use EasyVoice~\cite{Condado:07b} \cite{Condado:08a} \cite{Condado:09} as an example of a synchronous CMC application. The reason for using EasyVoice is because it is a good example to illustrate the need of more specific evaluation methods. Notice however that the concepts to be described in the paper are general and applicable to any kind of synchronous CMC,
and we indeed show the benefits of this test on regular instant messaging applications.

The remainder of the paper is organized as follows. Section~\ref{sec:user_requirements} presents user requirements elicited with a group of individuals with motor and speech disabilities. Section~\ref{sec:easyvoice} reviews EasyVoice, a system that allows people with voice disabilities to make phone calls using a synthesized voice, along with usability tests that have been done during its development. Section~\ref{sec:broader_evaluation} points to the limitations of those tests and suggests a more general evaluation procedure. Section~\ref{sec:turing} proposes an operational test for synchronous CMC that is based on one participant trying to guess whether another participant has a disability or not. 
Section~\ref{sec:test-protocol} describes the experimental design protocol that was made to validate our proposal, and Section~\ref{sec:test-results} presents and discusses the results obtained. Finally, Section~\ref{sec:conclusions} concludes the paper and presents future research directions.

\section{Eliciting User Requirements}
\label{sec:user_requirements}

We are concerned with designing mechanisms that allow people with motor and speech disabilities to communicate in a more effective manner. 
In order to better understand their requirements, we designed interviews using Activity Theory (AT) \cite{Kaptelinin:06}. AT provides a framework for describing the overall context of user activities \cite{Nardi:96}.  In AT the unit of analysis is the activity, which is defined in terms of the following components:
\begin{itemize}
\item activity participants and their motives (which emerge from spiritual or material needs).
\item object of the activity (driven by subject motives).
\item physical and symbolic instruments such as tools and signs (used to transform the object).
\item actions and operations (conscious and unconscious components of activities).
\item the community where the subjects belong and its underlying social rules.
\end{itemize}

Using AT raises awareness on the user activities in which CMC is used and how CMC could appropriately influence all its participants, and allows describing why and how people with motor and speech impairments interact with CMC in various situations.
AT lead us to learn more about their needs and motives, their current lifestyle, the way they use CMC applications in their daily life, the prejudices they encounter, the limitations they find, and also about their desires with respect to technology (i.e., features that they would like to see in CMC applications and other technological artifacts, that would be useful to them) as a means to perform their daily activities.

We interviewed five adult individuals with cerebral palsy, with an age range from 20 to 50 years old and varying levels of formal education. They exhibit varying degrees of motor disabilities but all are able to use regular keyboards, although they type slowly. With respect to speech, two of the individuals have moderate disabilities; their speech is only well understood for those you are used to talking to them. The other three individuals only have a mild speech disability, having no problem in being understood. All five individuals use e-mail, telephone, cell phone, and instant messaging, both for work and social activities. 

The two individuals with moderate speech disabilities believe that many times people are reluctant to say that they do not understand them and typically pretend they do. They find these situations annoying. The main problem, however, occurs in remote conversations (e.g., via phone or IP voice system) with people they do not know because they frequently hang up their calls. They think this happens because people think they are being teased, or due to lack of patience to make an effort to understand them. Because of this, they only use the phone to talk to people they know. When asked about their wishes with respect to CMC, both mentioned that they would like to have a technology capable of making their speech disability unnoticed so that they could use it for specific situations: to speak to people they do not know, or when having to ask for information or for services such as booking hotels. These are common tasks that they would like to perform as easily and as fast as people without speech disabilities, and would definitely like to avoid the effort, stress and frustration they feel when trying to communicate with people that do not understand them. In a nutshell, they just want things get done. Hence, the reason to make their disability unnoticed has nothing to do with shame but is rather for practical reasons. Neither of them want, or feel the need, to have their speech disability unnoticed for their social activities when they interact with friends or family (because members of the latter group either have no problem in understanding them, or are willing to make and effort to understand them). The other three individuals didn't express any special desire with respect to speech communication because their disability is so mild that they do not really have problems in  communicating verbally with other people.

With respect to typing, the five individuals said they would like to have a technology that would allow them to input text as easily and as quickly as people without disabilities. The reason in this case is also merely practical, they want to perform their tasks with less typing efforts and errors, which would also lead to less frustration, both for them and for the other people with whom they interact.

The next section reviews EasyVoice, a system developed in recent years that addresses, in part, some of the problems faced by some of these individuals. EasyVoice, however, has its own limitations, and is still far from the idealized CMC technology desired by this group of individuals.

\section{EasyVoice}
\label{sec:easyvoice}

\begin{figure*}
\center
\includegraphics[width=0.8\linewidth]{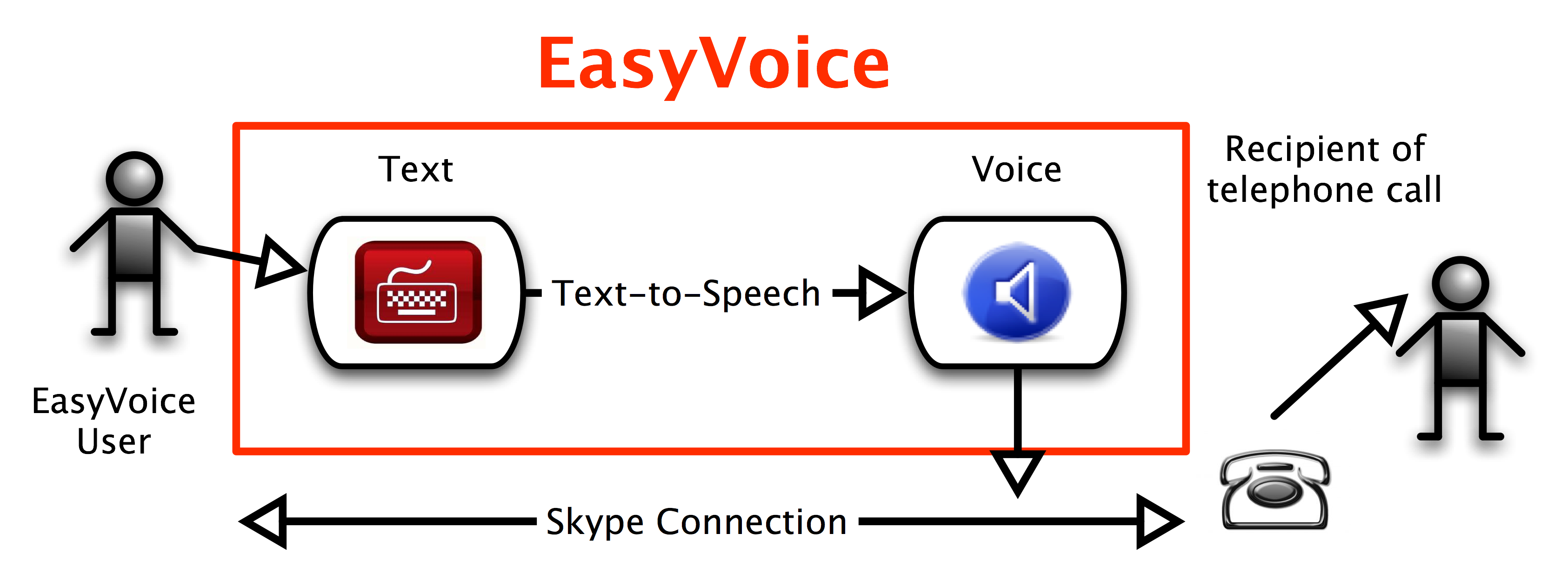}
\caption{Once a Skype connection is established, the EasyVoice user can type messages, the messages are converted into audio via text-to-speech, and the audio is injected into the Skype connection. At the other end of the conversation, the other person hears a synthesized voice. 
}
\label{fig:easyvoice}
\end{figure*}

EasyVoice is an application that integrates text-to-speech with voice over IP (VoIP) to allow people with voice disabilities to make phone calls using an artificial voice~\cite{Condado:07b} \cite{Condado:08a} \cite{Condado:09}. With EasyVoice, a person types text messages which are synthesized and sent through a computer network.
EasyVoice achieves this by working together with Skype via its Application
Programming Interface (API)\footnote{See~\url{https://developer.skype.com/}}.
Figure~\ref{fig:easyvoice} shows schematically how EasyVoice works. 

In addition to combining text-to-speech with VoIP, EasyVoice provides a number of features to accelerate the writing process. The design decision of including them was a result of the observation that many people with voice disabilities also happen to have motor disabilities. That's often the case of people with cerebral palsy.
The accessibility features provided by EasyVoice to accelerate the writing process are the following:
\begin{itemize}
\item word completion.
\item archive of recent messages.
\item abbreviation system.
\item optional virtual keyboard.
\end{itemize}

The word completion system searches in a built-in dictionary for
those words that have as a prefix the sequence of letters typed so
far by the user. 

The archive of recent messages is useful because
during a conversation it is many times necessary to repeat some
words or phrases, because the person at the other endpoint may not hear 
the sentence well enough. With the archive in
hand, the user does not need to retype the message again
and can simply pick it from the list of recently sent messages. 

Another important feature is the abbreviation system. It is common
for people to use abbreviations when writing. It is something that
is very popular in instant messaging software, especially among
young people. For example, in English it is common for people
to use ``{\tt btw}'' as an abbreviation of ``{\tt by the way}''. Within
EasyVoice, the user can define his/her own abbreviations.
The system will automatically
replace each abbreviation by the corresponding full spelled words,
before sending them to the speech synthesizer.

Finally there's an optional virtual keyboard with a scanning system incorporated 
which can be useful for those with more severe motor disabilities.

EasyVoice is an example that shows how assistive technology can allow people with disabilities to gain independence and to do things that they were unable to do before. At first sight it would seem that people with severe voice disabilities would be unable to benefit much from voice communication applications. Fortunately, EasyVoice shows that is not true. 
The next subsection presents a brief summary of the usability tests performed during the development of EasyVoice. 

\subsection{Usability tests done with EasyVoice}
\label{sec:easyvoice-tests}

The development of EasyVoice followed the principles of user-centered design and included an iterative design cycle with usability tests conducted at the end of each development phase. 

A group of people, all of them with voice disabilities and various degrees of motor coordination problems, were chosen to test the system at various stages of the development. The objectives of the tests were focused on the ability of users to write and send text messages, on the ease of use of the auxiliary features provided for accelerating the text input rate, on the simplicity with which the menu options could be chosen, and to assess how the EasyVoice features are used in a real conversation. The tests included both quantitative and qualitative measures. The quantitative ones measured the time needed to perform certain tasks (e.g, the time needed to type and send a given text message, the time needed to select a previously written message from the archive of recent conversations, and so on) and the number of errors made by the users while performing those tasks.
The qualitative measures were collected from a questionnaire where the users had the opportunity to express their level of satisfaction regarding the various features of the system. More details about the tests and its results are available elsewhere~\cite{Condado:09}. 

The EasyVoice tests were based on standard usability engineering practices that are advocated by HCI experts~\cite{Nielsen:94}. In particular the tests were used to assess the quality of the user interface in terms of how easy it is to use. In the next section, we show that these kind of tests, performed at one side of the communication alone, don't capture the full range of usability and user experience qualities that a synchronous CMC application should possess.

\section{An Accessibility-based Evaluation for Synchronous CMC}
\label{sec:broader_evaluation}

\begin{figure}
\center
\includegraphics[width=0.3\linewidth]{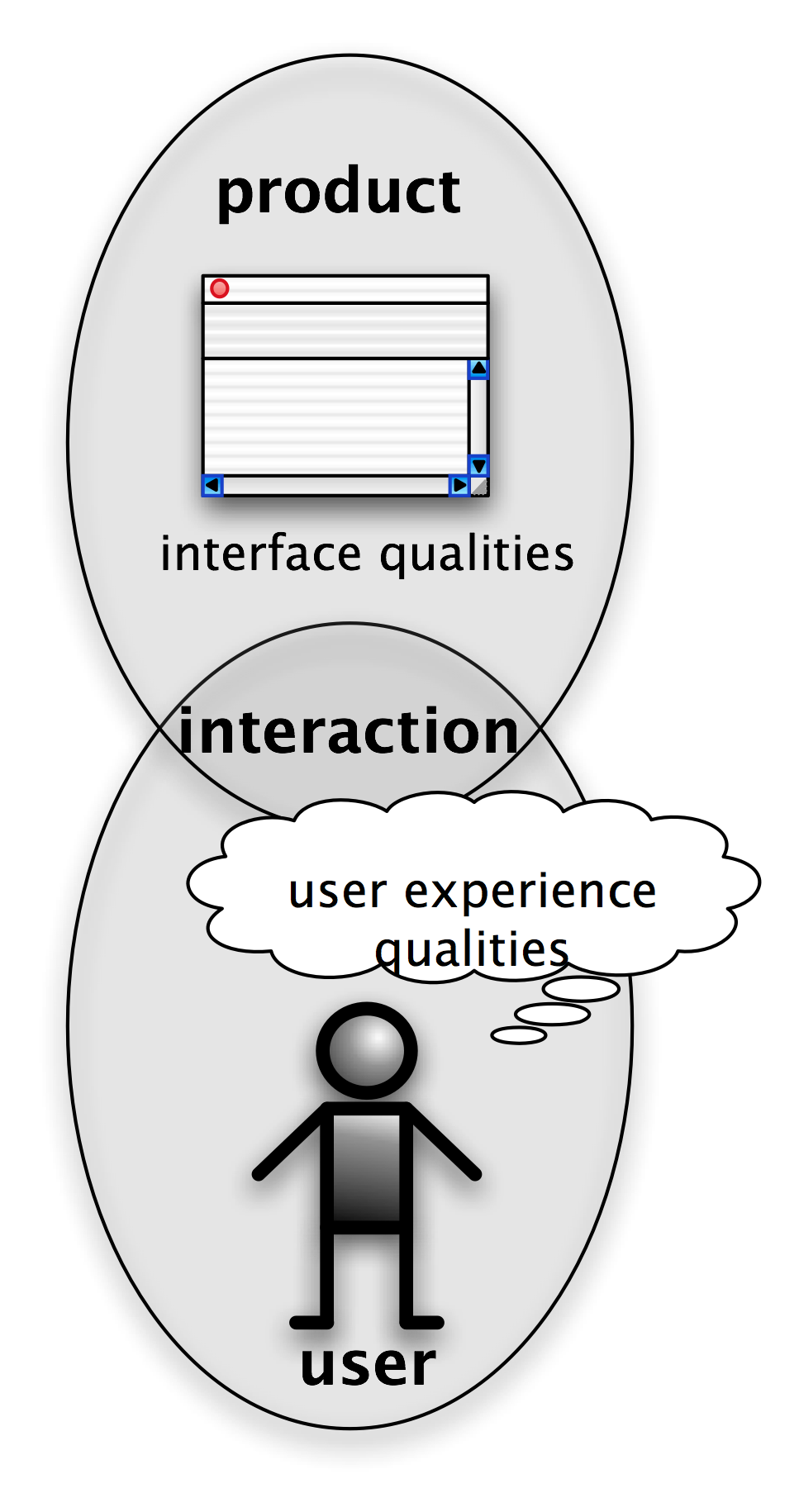}
\caption{Usability testing is used to assess the quality of the interaction between a user and a product.}
\label{fig:interaction}
\end{figure}

Interface and user experience qualities have been studied extensively in the HCI community, 
but have been mostly confined to asses the quality of the interaction between a user and a product, as depicted in Figure~\ref{fig:interaction}.
The usability tests with EasyVoice that were described in the previous section (and in more detail in \cite{Condado:09}) are a good illustration of this scenario. The tests measure how well the selected users were able to interact with the product (the EasyVoice interface in our case). 

One thing that stands out however is that this kind of single-sided tests are not sufficient to assess how good the interaction is. Being a synchronous CMC application, the evaluation should also take into account the user experience that occurs at the other endpoint of the conversation. 

The development of groupware applications, and in particular CMC applications, have expanded usability studies to include group and organizational levels of analysis. This focus shift led to the emergence of the CSCW field (Computer Supported Collaborative Work)~\cite{Olsen:03}. Hiltz and Johnson~\cite{Hiltz:90} propose dimensions to measure subjective satisfaction including two socio-emotional dimensions: (1) unexpressiveness (perceived inadequacy of the system for expressive, emotional, or personal communication) and (2) mode problems (perceived problems with computer-mediated communication). However, CSCW usability studies assume that participants have similar capabilities and thus, make no difference among senders and receivers, which results in defining a single evaluation instrument for them. Such symmetry is depicted in Figure~\ref{fig:interaction}.

In the specific context of EasyVoice, we should assess how good is the experience of the person who is talking to the EasyVoice user. (This observation has been made by Condado~\cite{Condado:09} but was not tested systematically.)
In other words, in the case of CMC applications where one of the participants has a disability, the interaction has a broader scope and goes beyond the typical scenario described in Figure~\ref{fig:interaction}.

\begin{figure*}[ht]
\center
\includegraphics[width=0.80\linewidth]{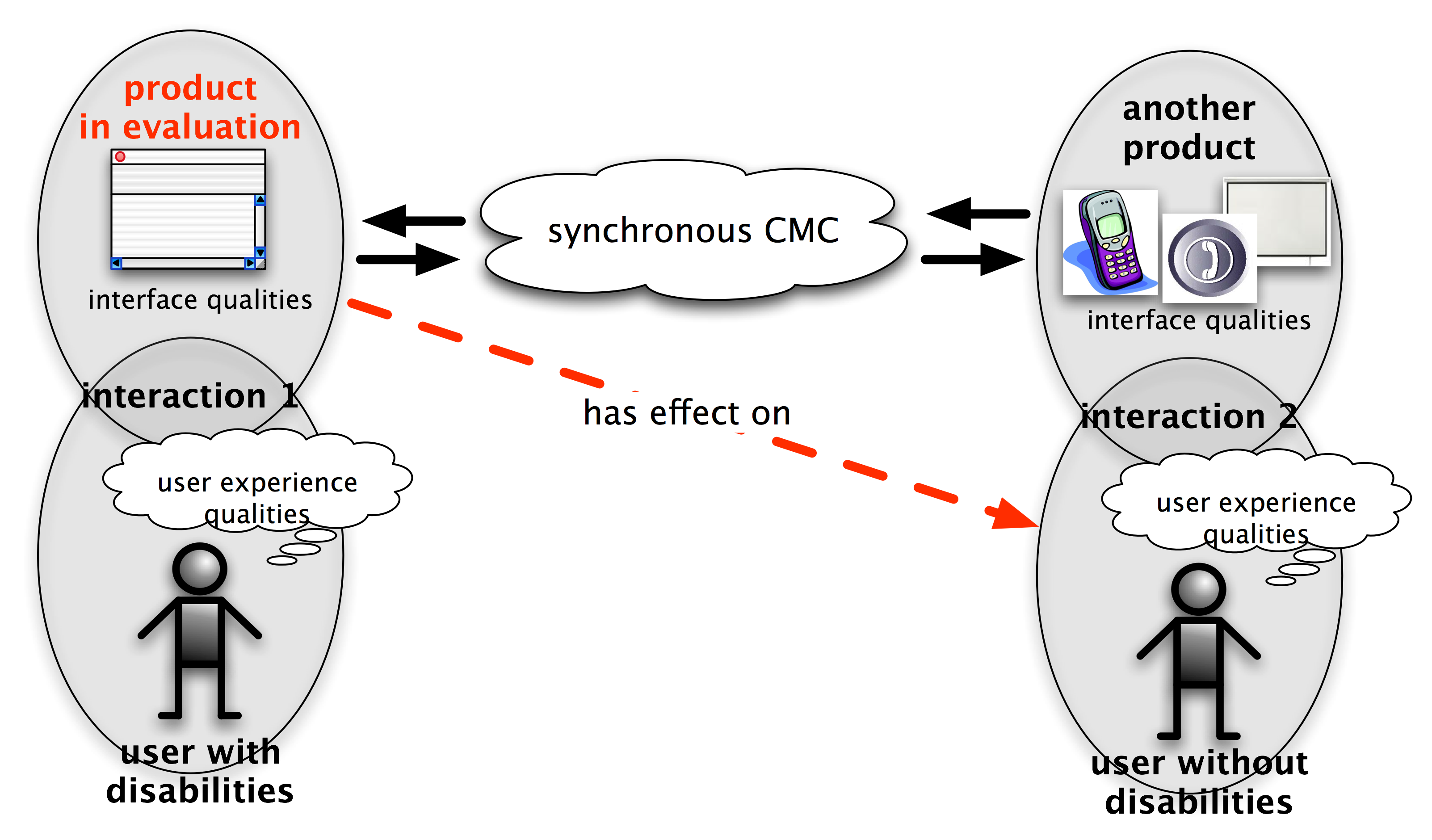}
\caption{The notion of interaction in a synchronous CMC application where one of the participants has a disability is broader. 
Each user interacts directly with a given product and these are described in the figure by 'interaction1' and 'interaction2'. 
It is important to notice  that the quality of each interaction, 1 and 2, results from interface and user experience qualities at each end party. 
However, the quality of 'interaction2' is also affected by the interface qualities of the product in evaluation.}
\label{fig:broader-interaction}
\end{figure*}

At this stage we would like to point out to one of the major limitations of the EasyVoice system, which is not captured by the kind of evaluation test conducted with a EasyVoice user: the delay that occurs in typing messages and subsequent text synthesis is an obstacle for a smooth voice conversation. While in text-based communication a small delay is considered acceptable, in voice-based communication such delays are usually not acceptable because participants in a voice conversation expect immediate feedback. This limitation of EasyVoice is not captured by a traditional evaluation procedure, because the problem itself is not confined to the the interaction between the EasyVoice user and the EasyVoice system; the delays that occur (due to the interaction of a user with the EasyVoice interface) have implications with respect to the user experience of the person that the particular user is talking to. 

For example, imagine a EasyVoice user with motor impairments who types messages very slowly, or imagine someone that interacts very slowly with the interface on an online multiplayer game due to motor coordination problems. This will likely cause frustration for the disabled player and boredom or even annoyance for the other players. This observation highlights that usability and user experience qualities such as efficiency and satisfaction need to measure not only how a given user interacts with the interface, but also the perceptions, emotions, and attitudes, of the other participants interacting with that user. This scenario is described schematically in Figure~\ref{fig:broader-interaction}  and contrasts with the traditional evaluation scenario (shown in Figure~\ref{fig:interaction}) where the communication is ``symmetric" in the sense that both end parties are likely to have a similar interaction behavior and their user interfaces are likely to be exactly the same.
To the best of our knowledge, this observation has not been made before in the context of evaluating synchronous CMC applications.

\section{The Disability Test}
\label{sec:turing} 

Our discussion so far suggests that an ideal synchronous CMC application should have an interface capable of making some of the user's disabilities unnoticed 
(and indeed this is supported by the wishes of the individuals we interviewed). 
Said differently, in an ideal case it would be good if participants cannot distinguish whether they are communicating with someone with disabilities. The last sentence suggests an evaluation metric that has striking similarities with the Turing Test.

The Turing Test  was proposed by Alan Turing as a replacement to the question ``Can machines think?" in a famous article published in 1950 ~\cite{Turing:50}. 
The Turing Test concept has evolved through time and its standard interpretation is usually a simplification of the original version.  Roughly speaking, the simplified version of the test consists of a computer being interrogated by a human via typewritten messages. The conversation can be on any imaginable subject. The computer passes the test if the interrogator cannot tell if there is a computer or a human at the other 
end~\cite{Russell:03}. 

Turing's ideas regarding this topic have been widely discussed and disputed in the fields of artificial intelligence, cognitive science, and philosophy of mind. Some considered Turing's paper to represent the beginning of AI and the Turing Test to be its ultimate goal, but others have attacked the idea and considered it to be useless~\cite{Saygin:00}. We are not going to delve into that discussion as it is irrelevant for our purpose here. But the idea behind the Turing Test brings up an important connection with respect to evaluating the user experience within the context of accessible synchronous CMC applications.

Again, for illustration purposes let us focus on the case of a voice-based conversation where one of the users (call him/her user A) is using EasyVoice, and the other user (call him/her user B) is holding a regular phone. Let's suppose that A has a motor disability and types very slowly. If during the conversation, B cannot tell whether A has a disability or not, then that's an evidence that A's interface provides good usability and user experience qualities for both A and B, at least in terms of efficiency and satisfaction.

Indeed, one could say that the ultimate goal when designing mechanisms for real-time interaction for people with disabilities is to guarantee that participants will not perceive the disabilities of the other participants. This is of course the ultimate goal, just like the Turing Test  has been suggested to be the ultimate goal of artificial intelligence. 

Having made this connection, we are now able to propose an operational test, which we refer to as 
the \emph{Disability Test}, to help to evaluate the user experience of a participant in a synchronous CMC application. Two variations of the test are proposed:

\begin{enumerate}
\item a binary test.
\item a test that involves a time component.
\end{enumerate}

The first one corresponds to the description that we have already made: either B thinks that A has a disability or not. In practice, an alternative to the binary test (yes/no answer) is to frame the question so that B can specify his level of agreement with the statement  ``A has a physical disability". In this case the possible answers could come from a Likert-based scale (for example 4 values: (1) definitely yes, (2) probably yes, (3) probably not, (4) definitely not). 

The second one records the moment, if any, when B is able to detect that A has a disability. Such moment can be measured quantitatively (e.g. absolute time) or qualitatively (e.g. beginning, middle, or end of the conversation).

The next section describes the design of an experiment to validate these ideas.

\section{Experimental Setup}
\label{sec:test-protocol}

To gather empirical evidence of the benefits of the proposed test, we designed an experiment 
and tested it on two examples of synchronous communication applications: (1) a Chat program
\footnote{Strictly speaking, Chat is a hybrid between synchronous and asynchronous CMC because it's usually acceptable if the participants don't respond to each other immediately. In our experiments however, we enforced the utilization of Chat as a synchronous communication application, as if the participants were holding an interview.} and (2) EasyVoice. 
The purpose of the experiment is not to compare Chat with EasyVoice, but rather to show how the 
proposed evaluation methodology can be used with two different case studies.

The experiment aimed at measuring usability and user experience qualities for both applications.
Such qualities were measured from a recipient's perspective (the recipient being someone that will try to guess whether the other participant has a disability or not). In order to achieve this goal, we defined the following scenario: A tourist wanting to visit 
a certain region
engages in a conversation over the Internet with two different persons, call them D and ND (D for disabled, ND for non-disabled), unknown to the tourist. The role of D and ND is to be tourist advisors by answering whatever questions the tourist might have. The conversation takes place one at a time, first the tourist talks with D and afterwards with ND, or vice-versa. 

Both advisors have similar knowledge of the selected region and one of them (D) has moderate motor disabilities. The same set of questions is made to D and ND (but different tourists can come up with different questions). In this experiment, we regard the tourist as the recipient and both usability and user experience qualities are measured on the recipient's side. In the case of Chat, the tourist and the advisor communicate with each other by typing text messages. In the case of EasyVoice, the tourist talks with his/her own voice and the advisor types messages which are automatically synthesized and sent via Skype. 
The recipient has no idea if D or ND has a disability or not. For the recipient, all scenarios are possible (nobody has a disability, both have a disability, or only one of them has a disability).

In terms of usability qualities, we measured the time in seconds that the recipient had to wait for each answer, as well as the length in characters of each answer. Such values allow a normalized metric of the speed in answering to the recipient's questions.
In terms of user experience qualities at the recipient's side, we measured the following ones through a questionnaire:
\begin{enumerate}
\item level of satisfaction with respect to the time needed to get answers.
\item motivation and  concentration during the conversation
\item opinion about the existence of some kind of motor disability from the person at the other endpoint of the conversation. 
\end{enumerate}
These questions were answered separately for D and ND. Questions in (1) and (3) where answered through Likert-based scales of four values. Questions in (2) where answered through yes/no values.

The tests were conducted by 20 subjects of similar age (in the range of 18 to 24 years old), background (university students), and a gender balance of roughly half male half female. All the subjects were used to communicate via instant messaging and none of them had previous experience in interacting with people with motor disabilities via CMC. The subjects were divided into 2 groups, where 10 interacted with D and ND using Skype's chat facility, and the other 10 interacted with D and ND using EasyVoice. The regular Chat conversations had a 3 minute duration and the EasyVoice conversations had a 5 minute duration. 

\section{Experimental Results}
\label{sec:test-results}

This section presents the results obtained in the experiment described above. We start by presenting a summary of the basic descriptive statistics of the results obtained for D and ND, for each tool. Then we present some tests that were done to assess the statistical significance of the results, and we finish with a discussion of the results.

\subsection{Descriptive Statistics}

Each participant (tourist) asked 4 to 5 questions in each Chat conversation totaling 86 questions where 39 were asked to D and 47 to ND. In EasyVoice conversations, each participant asked from 6 to 8 questions totaling 136 questions, 59 asked to D and 77 to ND.

\subsubsection{Speed in obtaining answers}

The mean and standard deviation of the speed in obtaining answers, measured in characters per second, is shown in Table~\ref{tab:speed}.

\begin{table}[h]
\center
\caption{Statistical values of the speed in obtaining answers from D and ND, measured in characters per second.}
\begin{tabular}{c|c|c||c|c|}
\cline{2-5}
& \multicolumn{2}{|c||}{Chat} & \multicolumn{2}{|c|}{EasyVoice} \\
\cline{2-5}
& advisor D & advisor ND & advisor D & advisor ND \\
\cline{1-5}
\multicolumn{1}{|c|}{Mean} & 0.97 & 2.93 & 1.23 & 1.98 \\
\multicolumn{1}{|c|}{Std Dev} & 0.24 & 0.75 & 0.65 & 0.51 \\
\cline{1-5}
\end{tabular}
\label{tab:speed}
\end{table}

The results for Chat and EasyVoice conversations show a clear difference in speed between D and ND, which is consistent with D's motor disability. However, the difference in speed between D and ND in EasyVoice conversations is smaller.

It is important to notice that with EasyVoice conversations there's an additional slowdown factor which is related to the time needed to do the text-to-speech conversion and the delivery of  the synthesized voice through the Internet. This explains the lower speed in ND's answers. Notice however that the speed of D improves substantially with EasyVoice (even with the text-to-speech slowdown factor) due to the typing aids provided by the application. This is an interesting result because it shows that the accessibility features available in EasyVoice help to attenuate the differences in typing speed between D and ND.

\subsubsection{Satisfaction with the speed of the answers}

\begin{figure}[h]
\center
\subfigure[Chat]{\includegraphics[width=0.45\linewidth]{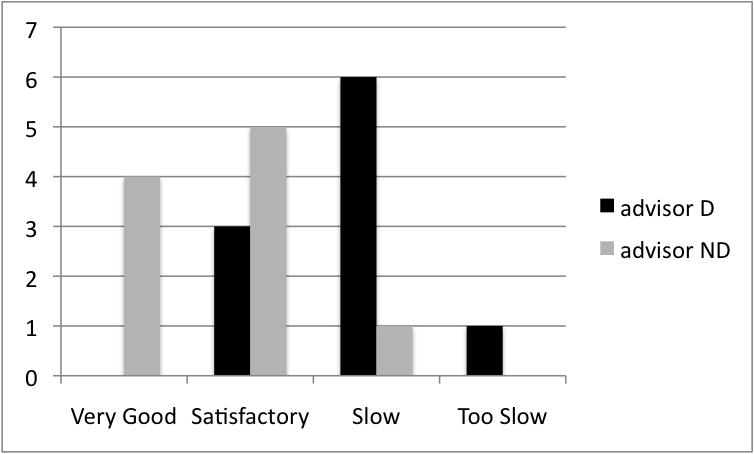}} \quad
\subfigure[EasyVoice]{\includegraphics[width=0.45\linewidth]{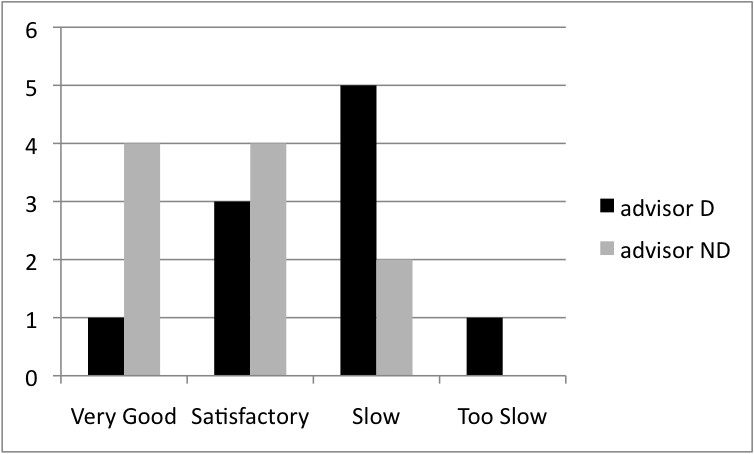}}
\caption{Recipients' level of satisfaction regarding the speed of the answers of D and ND.}
\label{fig:speed-of-answers}
\end{figure}

Figure~\ref{fig:speed-of-answers}-(a) compares the recipients' satisfaction regarding the speed of the answers of D and ND during the Chat conversations.  The results obtained for ND tend to be roughly a Likert point scale above from those obtained for D. 
The number of participants that found the speed of answers satisfactory is similar. However, whereas when interacting with D most of the participants found the speed of the answers to be slow, when interacting with ND most of them found it satisfactory. Very few participants found the speed of ND to be slow, and a significant number found the speed to be very good. When interacting with D, none of the participants found the speed of the answers very good. When interacting with ND none of them found the speed of the answers excessively low. 
The results for the EasyVoice conversations are similar (see Figure~\ref{fig:speed-of-answers}-(b)), but the differences between D and ND are less obvious.

\subsubsection{Motivation, concentration and distraction}

Table~\ref{tab:motivation} presents the results obtained with respect to whether the recipients kept motivated during the conversation or not. Table~ \ref{tab:distraction} presents similar results regarding whether the recipients lost their concentration and/or got distracted.

\begin{table}[h]
\center
\caption{Frequency table for the recipients' yes/no answers with respect to being motivated.}
\begin{tabular}{c|c|c||c|c|}
\cline{2-5}
& \multicolumn{2}{|c||}{Chat} & \multicolumn{2}{|c|}{EasyVoice} \\
\cline{2-5}
& advisor D & advisor ND & advisor D & advisor ND \\
\cline{1-5}
\multicolumn{1}{|c|}{Yes} & 4 & 9 & 5 & 9 \\
\multicolumn{1}{|c|}{No}   & 6 & 1 & 5 & 1 \\
\cline{1-5}
\end{tabular}
\label{tab:motivation}
\end{table}

\begin{table}[h]
\center
\caption{Frequency table for the recipients' yes/no answers with respect to getting distracted.}
\begin{tabular}{c|c|c||c|c|}
\cline{2-5}
& \multicolumn{2}{|c||}{Chat} & \multicolumn{2}{|c|}{EasyVoice} \\
\cline{2-5}
& advisor D & advisor ND & advisor D & advisor ND \\
\cline{1-5}
\multicolumn{1}{|c|}{Yes} & 6 & 2 & 4 & 2 \\
\multicolumn{1}{|c|}{No}   & 4 & 8 & 6 & 8 \\
\cline{1-5}
\end{tabular}
\label{tab:distraction}
\end{table}

With respect to keeping motivated during the conversation, the results obtained for Chat and EasyVoice were similar. 
A larger number of participants felt more motivated when interacting with ND than with D. In a follow-up question of the reason for loosing their motivation, all participants related it to the slowness of the answers.  
Similar results were obtained with respect to the lack of concentration and distraction.

\subsubsection{Opinion regarding the existence of a disability}

\begin{figure}[h]
\center
\subfigure[Chat]{\includegraphics[width=0.45\linewidth]{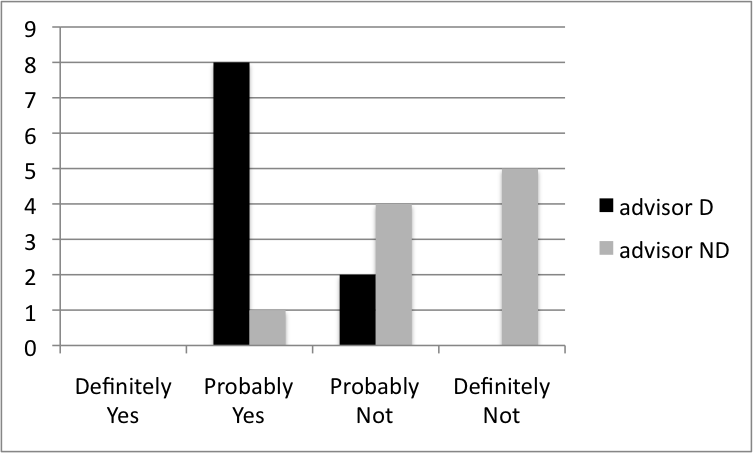}} \quad
\subfigure[EasyVoice]{\includegraphics[width=0.45\linewidth]{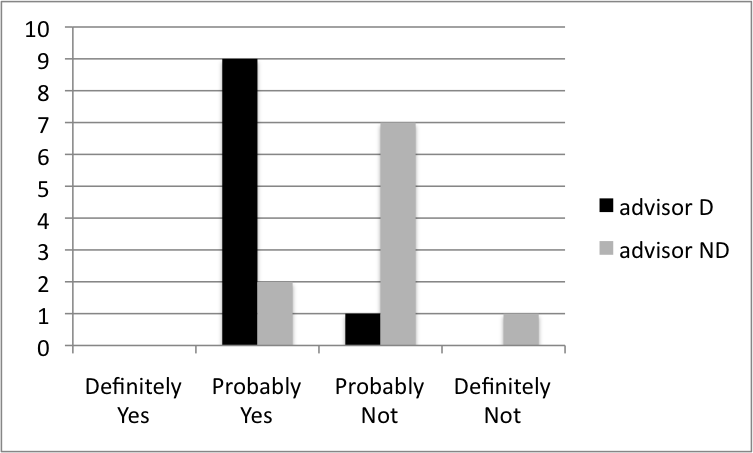}}
\caption{Recipients' opinion regarding the existence of a motor disability.}
\label{fig:has-disability}
\end{figure}

Figure~\ref{fig:has-disability} shows the results regarding the recipients' opinion of the possibility of D or ND having a motor disability. While none of the recipients exhibited an absolute certainty regarding the existence of a disability, it can be observed a clear difference between D and ND in the recipients' answers for both applications. According to the their answers, the chance of D having a disability is significantly greater than the chance of ND having a disability. Likewise, the chance of ND not having a disability is greater than the chance of D not having a disability. 

A relevant difference between Chat and EasyVoice was observed in a follow-up question about the moment of making such assumption. 
This corresponds to the second variation of the proposed Disability Test (see section~\ref{sec:turing}).
In the Chat case most recipients mentioned that they were able to infer the existence of a disability after the first two questions, while in the EasyVoice case most recipients were only able to detect that on the last questions. 

The results obtained show a clear difference between D and ND in both applications that are consistent with the actual speed of the answers shown in Table~\ref{tab:speed} and the fact that D has a motor disability. The difference in the speed of D observed in EasyVoice conversations is due to the typing facilities included in EasyVoice, which were not used by ND (probably because ND thinks they are unnecessary for him/her). These results confirm that EasyVoice indeed improves the typing speed of people with moderate motor disabilities.
However, while an apparent difference in terms of satisfaction regarding the speed of the answers of D can be observed between Chat and EasyVoice, such difference is less obvious in the other questions of the questionnaire.

\subsection{Statistical tests}

In order to evaluate if the apparent differences between D and ND have statistical significance, we proceeded to apply the following statistical tests:

\begin{description}
\item[Test 1:] Is there a significant difference in the answering speeds of D and ND?
\item[Test 2:] Is there a significant difference in the recipients' satisfaction level regarding the answering speed of D and ND?
\item[Test 3:] Is there a significant difference in the recipients' opinion regarding the possibility of D and ND having a disability?
\end{description}

\noindent
Test 1 was accomplished with Student's $t$-test for independent samples. 
The remaining tests were accomplished with the Freeman-Halton extension of the Fisher exact probability test for a two-row by four-column contingency table, suited for small expected frequency values ($<5$). The samples used in these tests were provided by the answers of the 20 participants that played role of the recipient (tourist) in each conversation.  

\subsubsection*{Test 1}

The null hypothesis for this test was: ``There is no difference between the answering speed of D and ND".  
For Chat conversations, the application of Student's $t$-test to D and ND answer speed samples, yields a two-tail $p$ value $< 0.0001$. For EasyVoice conversations, a $p<0.0001$ value was also obtained. 

Therefore, the null hypothesis can be rejected and it can be concluded that the differences in speed observed between D and ND, both in Chat and EasyVoice conversations, are statistically significant with a confidence level of 99\%.

\subsubsection*{Test 2}

The null hypothesis for this test was: ``There is no difference in the recipients' satisfaction level with respect to the answering speed of D and ND". For Chat conversations, the application of the Fisher test to the contingency table (distribution of satisfaction levels with D and ND) yields a $p$ value of $0.02$. The same test applied for EasyVoice conversations yields a $p$ value of $0.20$. 

Hence, the null hypothesis can be rejected in the first case (Chat conversations) with a confidence level of 95\% but not in the second (EasyVoice conversations). While we may conclude that there are 
significant differences in the recipient's satisfaction level with D and ND in Chat conversations, the same has no statistical significance for the case of EasyVoice conversations.
It is noteworthy that these results are consistent with the fact that the difference in speed between D and ND in Chat conversations is greater than in EasyVoice conversations. 

\subsubsection*{Test 3}

The null hypothesis for this test was: ``There is no difference between the recipients' opinion regarding the existence of a motor disability in D and ND." The application of the Fisher test to the contingency table corresponding to Chat recipients yields a $p$ value of $0.003$.
The same test applied to EasyVoice recipients yields a $p$ value of $0.005$.

Hence, the null hypothesis can be rejected in both cases with a confidence level of 99\%, and it can be concluded that there are significant differences in the opinion of both Chat and EasyVoice recipients regarding the existence of a possible motor disability in D compared with ND. 

\subsection{Summary and Discussion}

The tests presented aimed at measuring both usability and user experience qualities from the recipient's perspective.
The actual answering speed of our test subjects (D and ND) was registered in order to assess its correlation with respect to the recipient's satisfaction-related qualities, as well as to the recipient's capability of detecting the existence of a motor disability on the test subjects.

A descriptive statistical analysis shows apparent differences between D (user with motor disability) and ND (user without motor disability) regarding their answering speed, as well as in the recipient's satisfaction-related qualities. Statistical tests were conducted in order to assess the significance of the observed differences in the data collected and provided mixed results. While some differences were statistically significant, other apparent differences were not. 

Although EasyVoice improves the speed of D when compared with Chat (a tool not specifically designed for people with motor disabilities), such improvement is not enough to substantially reduce the difference in the recipient's level of satisfaction. This is an important result that could be easily overlooked without an evaluation at the recipient's side. A single-sided evaluation alone would reveal that the typing aids provided by EasyVoice allow a person with moderate motor disabilities to type messages faster (when compared with a regular Chat application with no accessibility features), but would miss the fact that those improvements are still insufficient to give a good user experience for those at the other end of the communication line. 

Strictly speaking, neither Chat nor EasyVoice passes the Disability Test. Although EasyVoice provides features that help to accelerate the writing process, and in some sense closes the gap between those with and without motor disabilities, those features were not sufficient to make the disabilities become completely unnoticed. Looking at the recipient's opinion regarding the existence of a motor disability, either for the case of Chat or EasyVoice (see Figure~\ref{fig:has-disability}), it can be observed that most people think that D probably has a physical disability. This is an indication that both applications have a lot of potential for improvement in terms of accessibility.

The reader may wonder if the goal of making the disabilities of a user become unnoticed during communication should be a major goal when designing accessible CMC applications. An alternative
goal that comes to mind would be to raise awareness to a communication partner so that the conversation can take place with an increased level of patience. We recognize that in some cases this can be a good strategy, but in other cases it is not of much help. For example, during the elicitation of requirements conducted at the beginning of this work, several individuals mentioned that in many situations, the communication partner does not have the patience, or simply cannot afford the time to wait and try to understand the disabled person. In the face of this evidence, it seems reasonable that the goal of making the disabilities of users unnoticed during communication is a sensible goal, at least in certain situations, and should be evaluated.

\section{Conclusions and Outlook}
\label{sec:conclusions}

This paper highlights the importance of testing user experience qualities at  the recipient's side in accessible applications intended for synchronous online interactions. Accessible interfaces for synchronous CMC applications should be designed in a way that minimizes frustration for the user with disabilities, and also minimizes boredom or even annoyance for the other users. This observation highlights that usability and user experience qualities such as efficiency and satisfaction need to measure not only how a given user interacts with the interface, but also the perceptions, emotions, and attitudes, of the other participants interacting with that user.

A novel contribution of this paper is the proposal of a test  which consists of a user trying to guess whether he is communicating with someone with disabilities or not. The resulting test is useful not only for evaluating the efficacy of the interaction, but also as a way of setting an ambitious goal when designing artifacts with accessibility in mind. Herein we argue that an important goal is to idealize interaction mechanisms that are able to
make the user's disabilities unnoticed during the communication process in order to increase its efficiency. If the other user doesn't notice the 
disabilities of the person communicating with him/her (during conversation) it suggests that those
disabilities had no disturbing effect in the communication.

The idea of using the proposed test is the result of a process of requirement elicitation based on Activity Theory, which was conducted through a qualitative study composed by interviews with five individuals with cerebral palsy. In these interviews we learned that for specific situations and activities, using a technology that suppresses their disabilities during communication is highly desirable. It is noteworthy that such desire obeys to strictly pragmatical reasons.

We used the proposed \emph{Disability Test} in two examples of synchronous CMC applications. In both cases,  the applications did not pass the test, and we interpret that as an indication that there is room for improvement with respect to their accessibility. It is important to mention that the Disability Test should be seen as a complementary test and by no means replaces the traditional usability tests.

It is not difficult to envision other application domains that can benefit from the proposed methodology. One that comes to mind is the development of accessible multiplayer games due to the degree of real-time interactivity that they often impose. The evaluation methodology advocated in this paper should be beneficial for designing such systems. 

The work presented in this paper has room for further developments and improvements. In particular, more extensive experiments should be performed and analyzed, with additional people taking the role of the advisors D and ND, and also with other types of CMC applications. Such experiments are likely to shed additional insights with respect to the utilization and interpretation of the proposed Disability Test. 
We are still far from our grand overall goal, which is to allow people with disabilities to communicate with others (with or without disabilities) in a truly effective manner. The evaluation methodology presented in this paper constitutes an important step in that direction.

\section{Acknowledgements}
We would like to thank all the people who participated in the evaluation tests.
Paulo Condado's work was sponsored by the Portuguese Foundation for Science and Technology, FCT/MCTES, under grant SFRH/BPD/65187/2009.

\bibliographystyle{plain}
\bibliography{references}

\end{document}